\documentclass[aps,amsmath,amssymb, twocolumn,
nofootinbib]{revtex4-1}

\usepackage{graphicx}
\usepackage{color}
\usepackage{bbold}

\renewcommand{\vec}[1]{{#1}}
\newcommand{\np}{N_p}
\newcommand{\nomp}{N_{1-p}}
\newcommand{\bit}{\begin{itemize}}
\newcommand{\eit}{\end{itemize}}
\newcommand{\tp}{\mathsf{T}}
 
\newcommand{\f}{\frac}
\renewcommand{\>}{\right\rangle}
\newcommand{\<}{\left\langle}
\newcommand{\ba}{\begin{align}}
\newcommand{\ea}{\end{align}}
\newcommand{\be}{\begin{equation}}
\newcommand{\ee}{\end{equation}}
\newcommand{\bi}{\begin{itemize}}
\newcommand{\ei}{\end{itemize}}
\newcommand{\lf}{\left(}
\newcommand{\ri}{\right)}

\newcommand{\Tr}{\operatorname{Tr}}
\newcommand{\tr}{\operatorname{tr}}
\newcommand{\str}{\operatorname{str}}

\begin{document}

\newcommand{\bra}[1]{\< #1 \right|}
\newcommand{\ket}[1]{\left| #1 \>}

\title{3D loop models and the $CP^{n-1}$ sigma model}
\author{Adam Nahum and J. T. Chalker}
\affiliation{Theoretical Physics, Oxford University, 1 Keble Road, Oxford OX1 3NP, United Kingdom}
\author{P. Serna, M. Ortu\~no and A. M. Somoza}
\affiliation{Departamento de F\'isica -- CIOyN, Universidad de Murcia, Murcia 30.071, Spain}
\date{26 July, 2011}

\begin{abstract}
Many statistical mechanics problems can be framed in terms of random curves; we consider a class of three-dimensional loop models that are prototypes for such ensembles.  The models show transitions  between phases with infinite loops and short-loop phases. We map them to $CP^{n-1}$ sigma models, where $n$ is the loop fugacity. Using Monte Carlo simulations, we find continuous transitions for $n=1,2,3$, and first order transitions for $n\geq 5$.  The results are relevant to line defects in random media, as well as to Anderson localization and  $(2+1)$-dimensional quantum magnets.\\

\end{abstract}
\maketitle

Loop models -- statistical mechanics problems whose degrees of freedom are loops or random walks -- are closely tied to field theory and more conventional statistical mechanics models, and give an alternative view on critical phenomena that has yielded new theoretical approaches and tools. More concretely, loops appear as topological defects, such as domain walls in two dimensions (2D) and vortices in three; as polymers; in the high-temperature expansions of lattice models, and in Monte-Carlo approaches to quantum problems. Questions about their statistics crop up in areas as diverse as Anderson localization \cite{glr, beamond cardy chalker}, turbulence  \cite{Bernard turbulence}, quantum chaos \cite{nodal lines}, cosmology \cite{cosmic strings}, optics \cite{dennis}, and frustrated magnetism \cite{jaubert}.

While there has been great progress in understanding 2D loop ensembles, the situation in 3D is less clear. This is not solely due to the unavailability of exact results: many qualitative questions are unanswered. Consider as an example random curves appearing in disordered media -- `deterministic walks in a random environment' (DWRE). The broad applicability of results for percolation cluster boundaries to 2D DWRE such as level lines of random height functions is well known, as are various continuum approaches to this problem \cite{duplantier saleur percolation, SLE reviews, read saleur}. But analogous 3D problems, such as the statistics of vortex lines in random fields, are not as well understood. Numerous problems of this kind have been simulated \cite{cosmic strings, dennis, tricolour percolation}, but it has not been clear which are in the same universality class, or what the relevant field theories should be.

In this paper we consider a family of three-dimensional loop models which are interesting from several points of view. A special case has appeared in the study of Anderson localization in 3D \cite{glr, beamond cardy chalker, ortuno somoza chalker, cardy class C review, cardy general graphs}, and results also apply to DWRE such as vortices. The loop models are also related to (2+1)D quantum magnets. They show transitions between short-loop phases and Brownian phases in which walks can escape to infinity. At critical points the loops have a nontrivial fractal structure.

Our aim is to understand the continuum descriptions of these models. We give an analytical mapping to lattice problems with more conventional (local) degrees of freedom. Coarse-graining then yields (compact) $CP^{n-1}$ models or supersymmetric variants, field theories which in two dimensions have been related to loop models by Read and Saleur \cite{read saleur}; see also \cite{candu et al, jacobsen et al, wolff strong coupling}. 
In addition, we perform Monte-Carlo simulations of the loop models, obtaining their phase diagrams and accurate values for critical exponents. These results support the identification of the continuum theory, since we find the expected exponents in the case (the $CP^1$ model) where they are known. For the $CP^2$ model we find exponents apparently for the first time. 

An important distinction is between oriented and unoriented loop ensembles.  While we focus mainly on the former, we argue that un-oriented loops are described by $RP^{n-1}$ models. Separately, an extension of the present work shows the general applicability of the $CP^{n-1}$ and $RP^{n-1}$ sigma models in the limit $n\rightarrow 1$ to problems of random curves in 3D disordered media \cite{forthcoming DWRE}.

\begin{figure}[b] 
\centering
\includegraphics[width=2.7in]{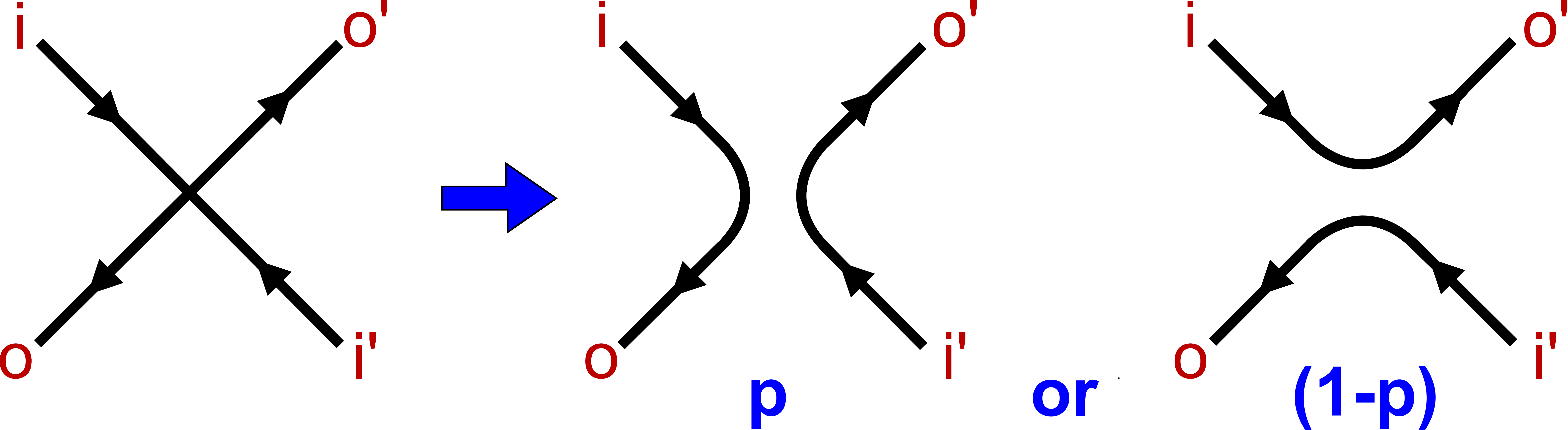}
\caption{Pairings at a node (with associated weights), and the labelling of links used in (\ref{lattice CPn-1}).}
\label{onenode}
\end{figure}

\emph{Loop models.} We consider loop models defined on four-coordinated, directed lattices, with two directed links entering and two leaving each node. A configuration $\mathcal{C}$ of completely packed, oriented loops is generated by pairing up the incoming and outgoing links at each node in one of the two ways compatible with their orientations (Fig.~\ref{onenode}). At each node, one of these pairings occurs with Boltzmann weight $p$ and the other weight $1-p$, with $0\leq p \leq 1$; the assignments are part of the definition of the model, along with the choice of lattice. We also give the loops a fugacity $n$. Let $|\mathcal{C}|$ be the total number of loops, and $N_p$ the number of nodes where the weight-$p$ pairing is followed. The partition function is
\be
Z_{\text{loops}} = \sum_\mathcal{C} p^{N_p} (1-p)^{N_{1-p}} n^{|\mathcal{C}|}.
\ee
When the fugacity is a positive integer, it can be reproduced by a sum over $n$ `colours' for each loop. 
 The models with $n=1$ have the property $Z_{\text{loops}}=1$: In this case, the node configurations are independent random variables, and the walks are DWRE. In general, as $n$ does not flow under renormalization \cite{Kondev}, we think of it as labelling different models, and of $p$ as a parameter.

\begin{figure}[t] 
\centering
\includegraphics[width=3in]{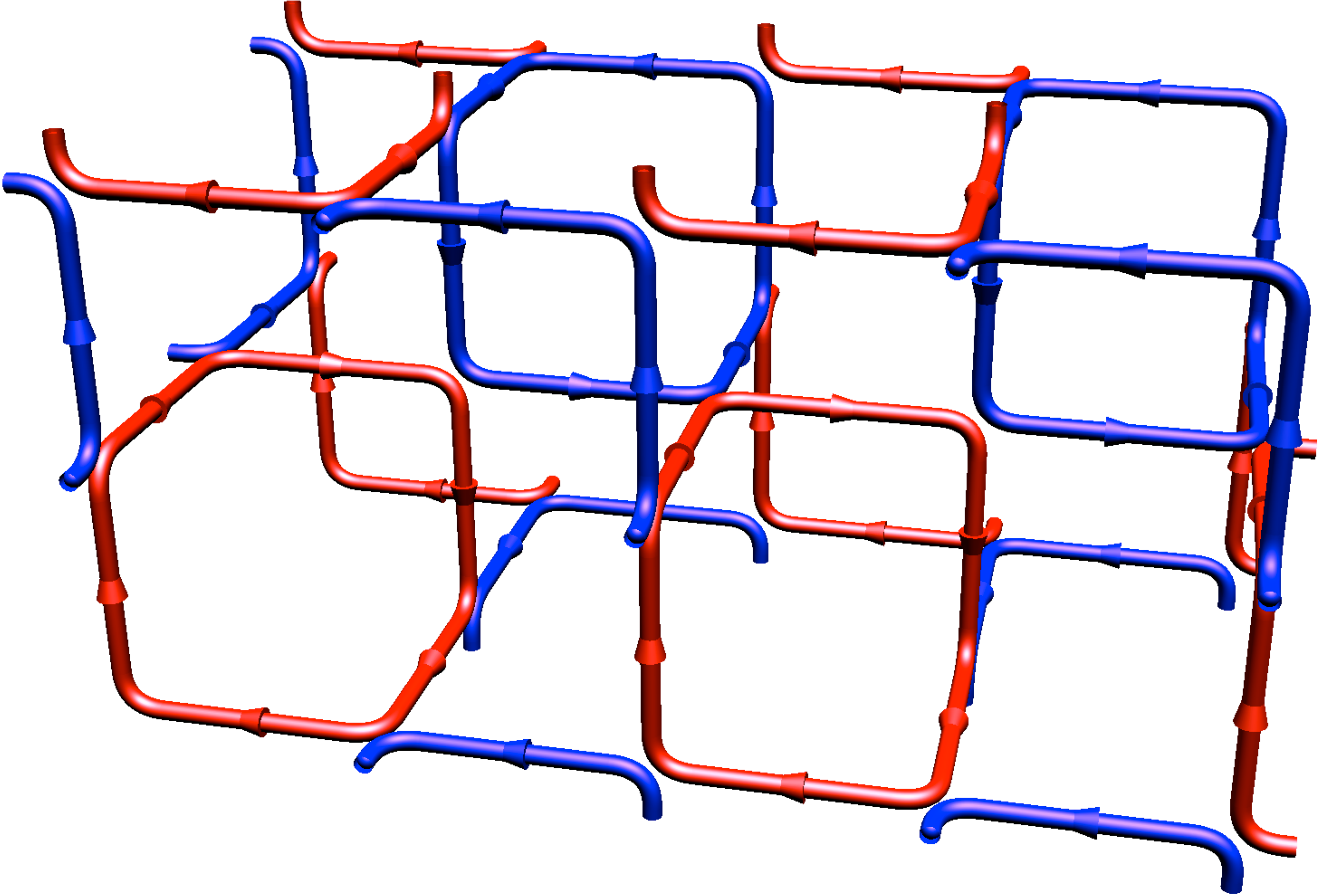}

\caption{Loops on the 3D K-lattice at $p=0$. At $p=1$, they become infinite straight trajectories (crossing at nodes).}
\label{3D K lattice}
\end{figure}

This general  recipe can be used to construct various models, depending on the lattice and node assignments. Here we are interested in models with transitions between localized and extended phases, and in the universal behaviour at the transition and in the extended phase. Our simulations use Cardy's `3D L-lattice' \cite{cardy class C review}, which has cubic symmetry, and a variant, the  `3D K-lattice', 
 which differs in its link orientations and phase diagram (Fig.~\ref{3D K lattice}).


\emph{Lattice magnet}. We rewrite $Z_{\text{loops}}$ in terms of local `magnetic' degrees of freedom which can be coarse-grained in a fairly straightforward way. In this we are inspired by the well-known $O(n)$ loop models \cite{domany} -- here we obtain instead a lattice $CP^{n-1}$ model.

Introduce complex unit vectors $\vec{z}_l=(z^1_l,...,z^n_l)$ on the links $l$ of the lattice, and denote the integral over these degrees of freedom by $\Tr$ (normalized so $\Tr 1 =1$). Now consider a Boltzmann weight which is a product of terms, one for each node of the lattice. Labelling the incoming and outgoing links at a given node as in Fig.~\ref{onenode}, 
\be
\label{lattice CPn-1}
Z = \Tr \prod_{\text{nodes}} \lf
p (\vec{z}_o^\dag \vec{z}_i)(\vec{z}_{o'}^\dag \vec{z}_{i'})+(1-p) (\vec{z}_{o}^\dag \vec{z}_{i'})(\vec{z}_{o'}^\dag \vec{z}_i)
\ri.
\ee
This partition function reproduces the sum over loops with the right weights. To see this, note that the terms in the expansion of the product over nodes are in correspondence with loop configurations $\mathcal{C}$:
\be
\label{Z intermediate}
Z = \Tr \sum_{\mathcal{C}} p^{\np}(1-p)^{\nomp}  
\prod_{\mathcal{L}\in\mathcal{C}} 
\tr \prod_{l\in \mathcal{L}} (z_l z_l^\dag).
\ee
Here $\mathcal{L}$ is a loop, the outer product $z_l z_l^\dag$ is an $n\times n$ matrix, and `$\tr$' 
is a trace in this space (the ordering of the last product is given by the sequence of links on $\mathcal{L}$).  Now, since $\Tr z_l z_l^\dag=\mathbb{1}/n$, we are left with one $n\times n$ trace, i.e. one `colour' index to sum, per loop.  Let $N_l$ be the total number of links on the lattice. Then
\be
\label{Z with colours}
Z = \f{1}{n^{N_l}}\sum_{\mathcal{C}}\sum_{\text{loop colours}} p^{\np}(1-p)^{\nomp}=\f{1}{n^{N_l}}Z_{\text{loops}}.
\ee
The Boltzmann weight (\ref{lattice CPn-1}) defines a classical magnet for the `spins' $z$. In addition to the unitary global symmetry, it has the local $U(1)$ symmetry $z_l\rightarrow e^{i\phi_l} z_l$, so our spins live not on the sphere $|z|^2 =1$ but on complex projective space, $CP^{n-1}$. This space degenerates to a point when $n=1$, leaving no degrees of freedom. Thus we must either resort to a replica-like limit $n\rightarrow 1$, or generalize (\ref{lattice CPn-1}) to a supersymmetric theory by replacing $z$ with a unit supervector of $n+k$ bosonic and $k$ fermionic components, $\psi=(z^1,...,z^{n+k},\chi^1,...,\chi^k)$. A nonzero number $k$ of fermions leaves the partition function and its loop representation unchanged (using $\Tr\psi\psi^\dag =\mathbb{1}/n$, the loop expansion goes through as before, with $\tr\rightarrow\str$) but yields more operators, and is necessary to give a nontrivial theory when $n=1$ (or $n<1$). 

\emph{Field theory.} The naive continuum limit of (\ref{lattice CPn-1}) is the $CP^{n-1}$ model. In a sigma model formulation, with an auxiliary gauge field $A$ to remove the unwanted phase degree of freedom, the Lagrangian density is:
\begin{equation}
\label{cpn with A soft}
\mathcal{L} = \f{1}{g^2} |(\partial - i A) \vec z|^2 \, ,
\quad {\rm with} \quad |z|^2=1\,.
\end{equation}
The SUSY version, the $CP^{n+k-1 | k}$ model,  is got by $z\rightarrow \psi$. A crucial point in any formulation is that the gauge field is \emph{compact}: the set of gauge transformations $\vec{z}\rightarrow e^{i\phi} \vec{z}$, $A\rightarrow A+\partial \phi$ is larger than in noncompact $U(1)$ gauge theory as $\phi$ can jump by $2\pi$. This implies that Dirac strings of flux $2\pi$ incur no cost in action, and that in integrating over $A$ we must include Dirac monopole configurations with quantized charge \cite{polyakov compact gauge theory}.

Work on deconfined criticality \cite{deconfined} has made clear that compactness is a subtle issue, so it is useful to have another route to the continuum limit for the loop models. We use the transfer matrix to extract a $(2+1)$D quantum $SU(n)$ antiferromagnet \cite{large n for square lattice AF, numerics for dimerization} as an intermediate step -- for an analogue in 2D, see \cite{read saleur, candu et al}. This procedure, to be described in \cite{forthcoming long version}, clarifies the compactness of $A$.

Finally, an alternative to (\ref{cpn with A soft}) is to use explicitly gauge-invariant degrees of freedom. The two-colour case $n=2$ reduces simply to the $O(3)$ (classical Heisenberg) model via $S^\mu = z^\dag \sigma^\mu z$, with $\sigma$ a Pauli matrix, and indeed the loop models with $n=2$ show the usual $O(3)$ exponents as described below. For general $n$ (without fermions) we can use the traceless matrix $Q = \vec{z}\vec{z}^\dag - 1/n$.

\emph{Correlators.} In 3D, the $CP^{n-1}$ model has a transition between a disordered phase and an ordered phase with $2(n-1)$ Goldstone modes (or $2[n+k-1]$ bosonic Goldstone modes and $k$ complex fermions). Translating correlators of gauge invariant operators into loop language shows that the former corresponds to the localized and the latter to the extended phase of the loop model. 

Consider $G_N (r)$, the probability that two small regions separated by a distance $r$ are connected by $N$ distinct strands of loop. In the localized phase and at a critical point all loops are finite, and $G_N(r)$ is non-zero only for even $N$. $G_2(r)$, the probability that two distant points lie on the same loop,  is proportional to $\<\tr Q(0) Q(r)\>$, and higher correlation functions $G_{2M}(r)$ can be written as two-point functions of operators $(z^1 \bar z^2)^M$. In the localized phase $Q$ is massive and $G_2(r) \sim r^{-1} e^{-r/\xi}$. The typical loop size $\xi$ diverges on approaching the critical point at $p=p_c$, as $\xi \sim |p-p_c|^{-\nu}$, and at criticality $G_2(r) \sim 1/r^{1+\eta}$, where $\eta$ is the anomalous dimension of the gauge invariant spin $Q$ (or the Heisenberg spin $S$ when $n=2$). A simple scaling argument \cite{scaling relations, duplantier saleur percolation} relates $\eta$ to the fractal dimension $d_f$ of the critical loops, and to the 
exponent $\tau$ governing the distribution, $P(l) \sim l^{1-\tau}$, of the length $l$ of the loop through a given link:
\ba
d_f &= \f{5-\eta}{2}, & \tau &= \f{11-\eta}{5-\eta}.
\end{align}
Since we expect $\eta$ to be small, $d_f$ will be close to $5/2$. Interestingly, the mean field value $\eta=0$ does not give the `trivial' fractal dimension of two -- this is due to confinement of $z$ into a gauge invariant composite field. Confinement also has an interpretation as a relation between different loop ensembles (essentially worldlines of $z$ versus worldlines of $S$) when $n=2$.

An important basic consequence of the $CP^{n-1}$ description is Brownian behaviour in the extended phase, which has been observed but not derived in related problems \cite{cosmic strings, dennis, jaubert, tricolour percolation, ortuno somoza chalker}. In this phase, contributions from finite strands of infinite walks make $G_N(r)$ non-zero for both even and odd $N$: $G_N(r) \sim r^{-N}$. These are Brownian exponents (each factor of $1/r$ is just the probability that a random walker visits a given site at distance $r$ from its origin) and imply a fractal dimension of two. They follow from free field theory for the Goldstone modes, allowing for the fact that both the properties of long loops and symmetry breaking in the $CP^{n-1}$ model are sensitive to boundary conditions. If walks can end on the boundary, `infinite' walks are those that do so: they have typical length of order $L^2$. With periodic BCs, all walks form closed loops: though Brownian, the `infinite' loops have typical length of $O(L^3)$, since a random walker trapped in a region of linear size $L$ will on average visit $O(L^3)$ sites before re-finding his starting point. The probability of a given link lying on an infinite loop is proportional to the order parameter, so varies as $|p-p_c|^\beta$ close to $p_c$.

\emph{Unoriented loops.} Models with {un}oriented loops are also interesting, e.g. in relation to polymers and ${\mathbb{Z}}_2$ vortices. Similar arguments relate them to $RP^{n-1}$ sigma models, with real spins. In the models considered above, allowing node pairings which do not respect the link orientation -- so loops have no well-defined orientation -- corresponds to a perturbation $\delta \mathcal{L}\propto - \tr Q^\tp Q$, favouring real $Q$s and causing a crossover to $RP^{n-1}$ behaviour.

\begin{figure}[t]
   \centering
   \includegraphics[height=1.252in]{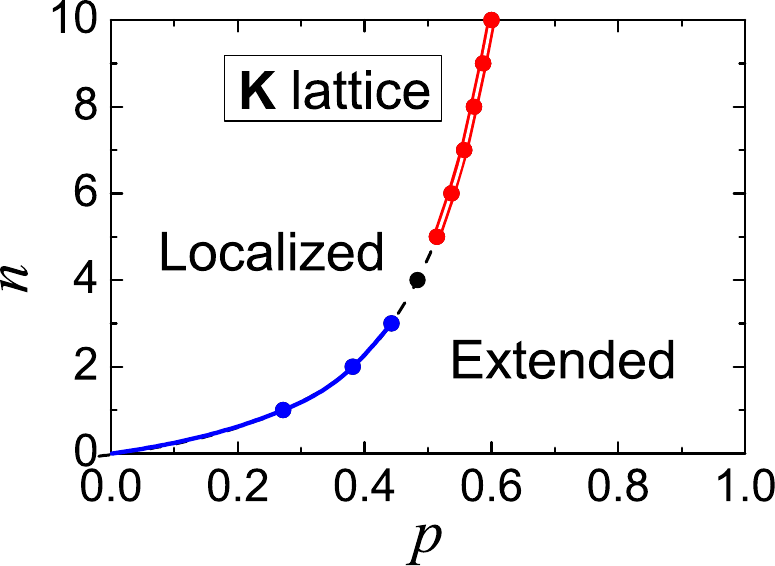}    \includegraphics[height=1.25in]{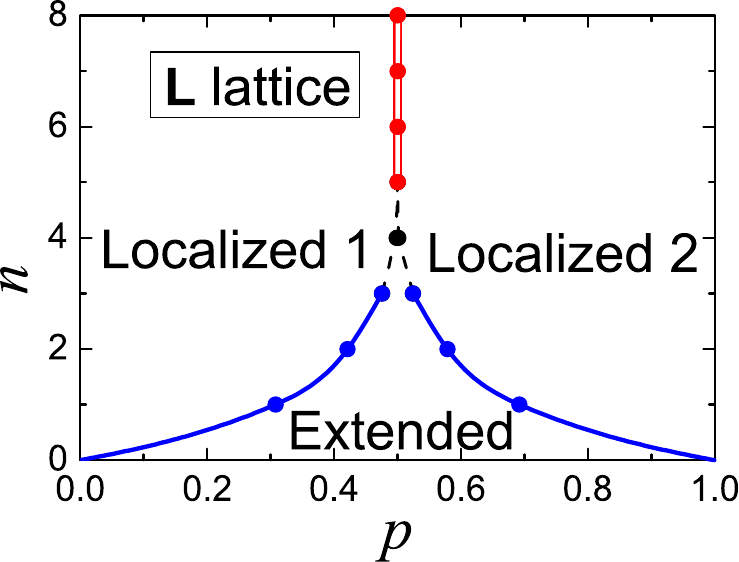} 
\caption{Phase diagrams for the K and L-lattices. Continuous transitions are indicated by blue dots and single line and first order transitions by red dots and double line.}
\label{phasediagrams}
\end{figure}
\emph{Deterministic walks in a random environment.} A key outcome of this work is the general applicability of the $CP^{k|k}$ model to oriented loops in short-range correlated random media -- most notably various kinds of vortices, such as optical vortices \cite{dennis}, cosmic strings \cite{cosmic strings},  XY vortices in the paramagnetic phase \cite{forthcoming DWRE}, and `tricords' in tricolour percolation \cite{tricolour percolation}.  The striking compatibility of the exponents in \cite{tricolour percolation} and in the $n=1$ loop model \cite{ortuno somoza chalker} confirms that these problems are in the same universality class. Derivations will appear separately \cite{forthcoming DWRE}.

\emph{Numerical results.} We use Monte Carlo simulations to study the loop model (1) on the 3D K and L-lattices introduced above, for integer $n$ in the range $1\leq n \leq 10$. We take samples of linear size $32\leq L \leq 100$ in units of the link length, with periodic BCs. The fugacity $n$ is introduced via  loop colours, and two types of elementary Monte Carlo move are employed: either a change in the colour of one loop, or a change in the configuration of a node whose links all carry the same colour. Typical run lengths involve of order $10^5$ Monte Carlo steps of each type, per loop or node respectively. 
For details, see \cite{forthcoming long version}.

\begin{figure}[b]
\centering
\includegraphics[height=2in]{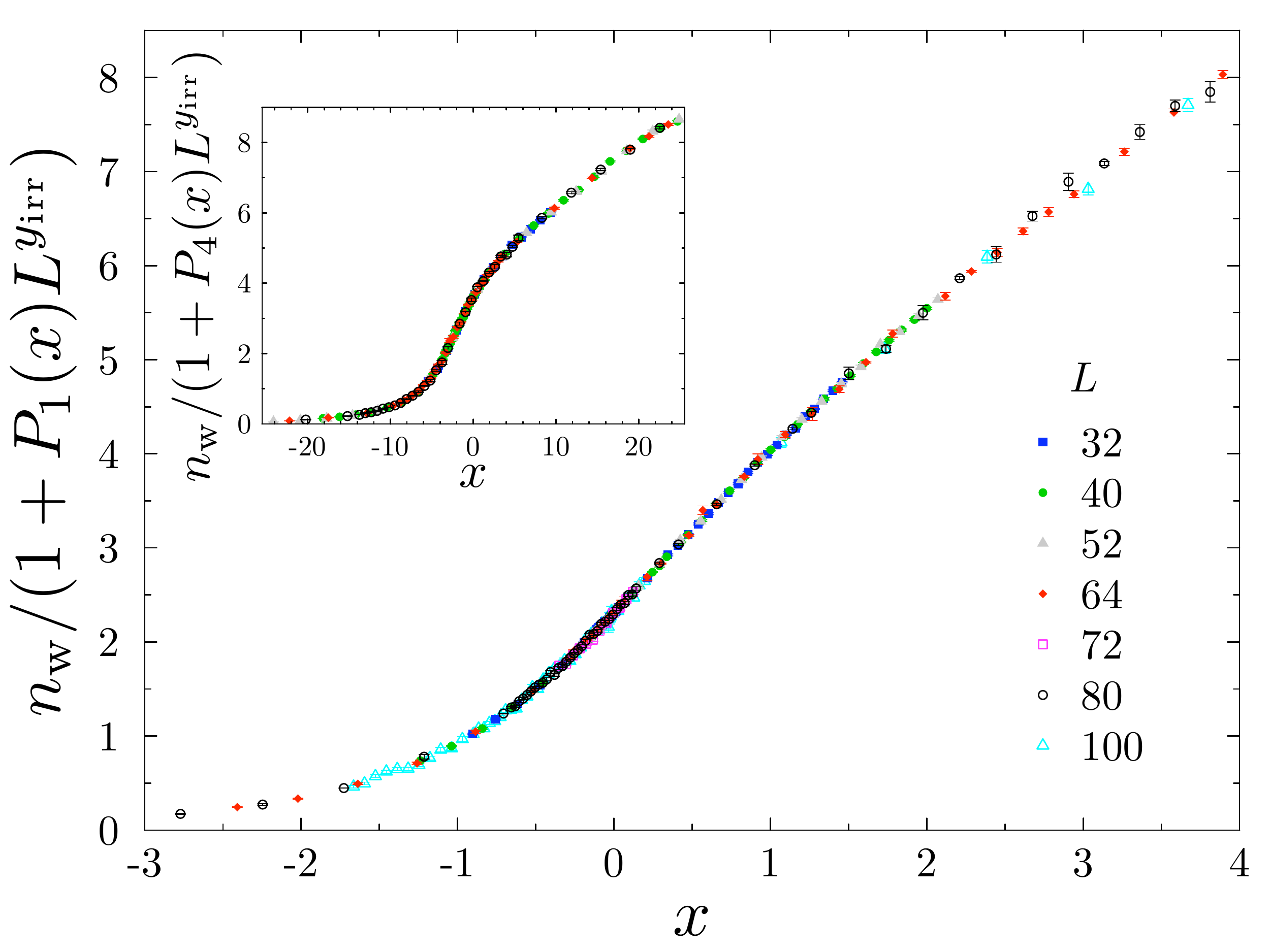} 
\caption{Scaling collapse for $n_{\rm w}(p,L)$ on the 3D K-lattice. Main panel: $n=2$; inset: $n=3$.}
\label{scaling collapse}
\end{figure}

Phase diagrams are shown in  Fig.~\ref{phasediagrams}. The 3D K-lattice is constructed to have only short loops at $p=0$ but infinite ballistic trajectories at $p=1$. Simulations show a single transition between a localized and an extended phase, which is continuous for $n\leq3$, and first order for $n\geq 5$ in agreement with a simple large $n$ treatment. 

The 3D L lattice is symmetric under $p\to 1-p$ and has only short loops at $p=0$ or 1. For $n\leq 3$ it has an extended phase around $p=1/2$, separated by continuous transitions from localized phases at large and small $p$. For $n\geq 5$ it has only localized phases, and a first order transition at $p=1/2$. Work is in progress to resolve behaviour at $n=4$ on both lattices. Previous  Monte Carlo studies of $CP^{n-1}$ \cite{duane green, kataoka} obtained a first order transition at $n=4$.

We present studies of critical behaviour for transitions on the K-lattice at $n=2$ and 3. Results for the 3D L-lattice
are consistent with universality when compared with the K-lattice at these values of $n$ and with previous work \cite{ortuno somoza chalker} at $n=1$. We examine two observables. One is the average number $n_{\rm w}(p,L)$ of curves spanning the sample in a given direction. The other is the susceptibility, which can be expressed \cite{forthcoming long version} in terms of the average number $n(l)$ of loops of length $l$, as $\chi(p,L) = L^{-3}\sum_l l^2n(l)$.

Empirically, the scaling of the winding number, including finite size corrections, is adequately described by the form $n_{\rm w}(p,L) = f(x)(1+P_m(x)L^{y_{\rm irr}})$, where: $x$ is the scaling variable, $x\simeq L^{1/\nu} (\delta p + A\delta p^2)$; $\delta p = (p-p_{\rm c})$; $\nu$ is the correlation length exponent; $y_{\rm irr} < 0$; $P_m(x)$ is a polynomial of order $m$; and $f(x)$ is constructed using splines. Results are shown in Fig.~\ref{scaling collapse}. We fit $\chi(p,L)$ in a similar manner, using the susceptibility exponent $\gamma$. Values for $y_{\rm irr}$ ($-1.0(3)$ and $-0.6(4)$, at $n=2$ and $3$ respectively) have large uncertainties, but those of $\nu$ and $\gamma$ are much more precise as finite size corrections are small.

\begin{figure}[t]
\centering
\includegraphics[height=2in]{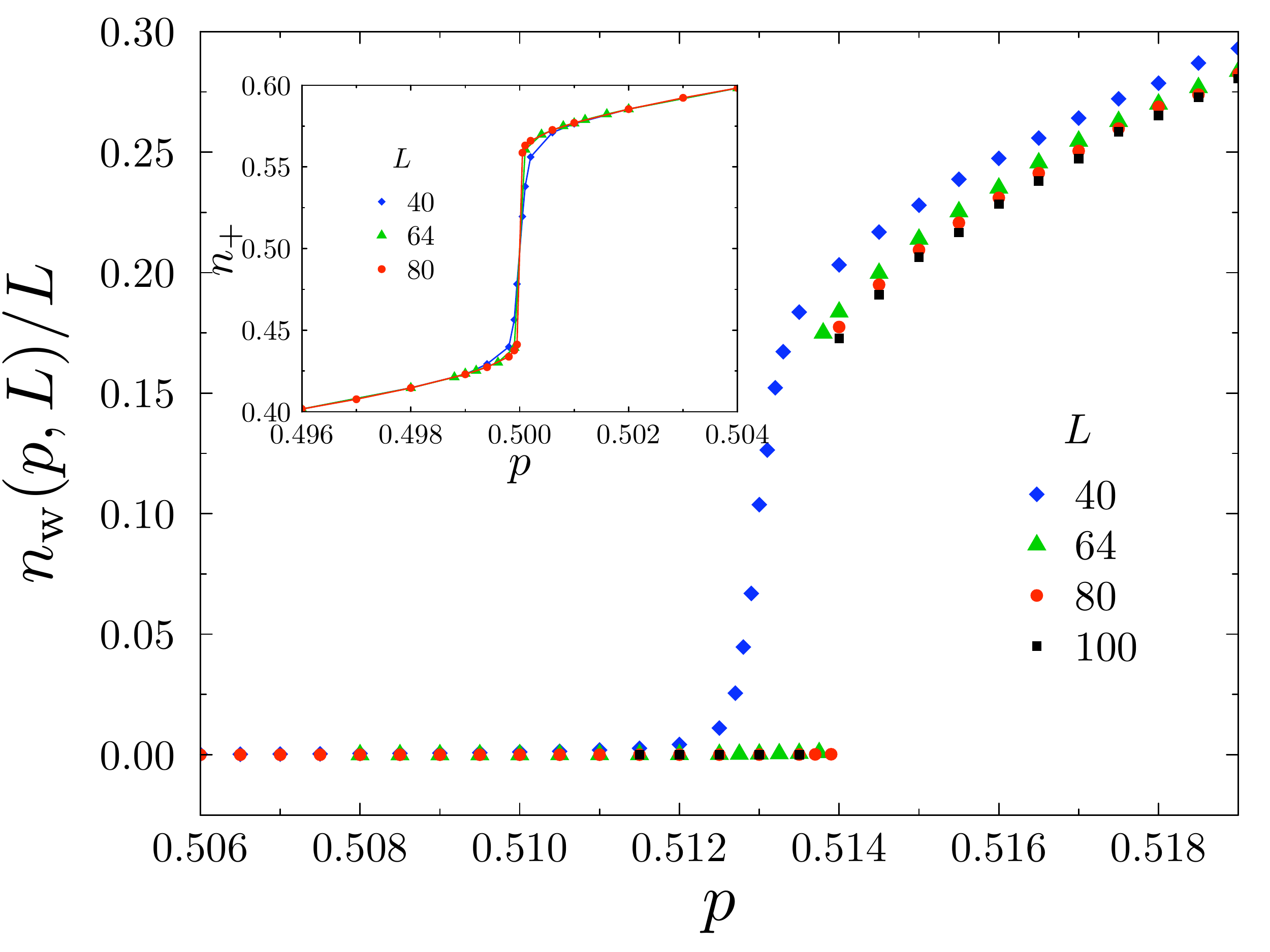} 
\caption{First order transitions at $n=5$: (main panel) the jump in $n_{\rm w}$ on the K-lattice and (inset) in $n_+$ on the 3D L-lattice.}
\label{first order}
\end{figure}

For $n=2$ we obtain $\nu=0.708(5)$ and $\gamma=1.39(1)$, fitting to over 300 data points. We believe that the consistency of these values with previous high-precision studies of the 3D classical Heisenberg model ($\nu=0.7112(5)$ and $\gamma=1.3960(9$) \cite{O(3) numerics}) provides compelling support for our identification of the loop model with $CP^{n-1}$. 
For $n=3$ we find $\nu=0.50(1)$ and $\gamma=1.01(2)$. We are not aware of a previous determination of exponents for $CP^2$.

In contrast, for $n\geq 5$ we find clear evidence of first order transitions on both lattices, as displayed in Fig.~\ref{first order}. For the K-lattice there is a rapid change in $n_{\rm w}(p,L)$ as $p$ passes through $p_{\rm c}$, developing into a step with increasing $L$. For the 3D L-lattice $n_{\rm w}(p,L)\to 0$ for large $L$ at all $p$. A transition at $p=1/2$ between distinct localized phases is signaled by a step in $n_+ = \langle N_p\rangle/(N_p + N_{1-p})$, which can be viewed as the internal energy density.

We thank E. Bettelheim, P. Fendley, I. Gruzberg, A. Ludwig, P. Wiegmann and especially J. Cardy for discussions. This work was supported by EPSRC Grant No.
EP/D050952/1, by DGI Grant Nos. FIS2009-13483 and AP2009-0668, and
by Fundacion Seneca, Grant No. 08832/PI/08.

\end{document}